\documentstyle[12pt]{article}
\begin{document}
\hfuzz=1pt
\setlength{\textheight}{8.5in}
\setlength{\topmargin}{0in}
\begin{center}
\Large {\bf  Bell's theorem, quantum mechanical non-locality and
atomic cascade photons} 
\\  \vspace{.75in}
\large {M. Ardehali}
\\ \vspace{.3in}
Research Laboratories,
NEC Corporation,\\
Sagamihara,
Kanagawa 229
Japan

\end{center}
\vspace{.20in}

\begin{abstract}
Bell's theorem of $1965$ is a proof that all realistic interpretations
of quantum mechanics must be non-local. 
Bell's theorem consists of two parts: first a correlation
inequality is derived
that must be satisfied by all local realistic theories; second
it is demonstrated
that quantum mechanical probabilities
violate this inequality in certain
cases.
In the case of ideal experiments,
Bell's theorem has been proven. 
However, in the case of real experiments where polarizers and
detectors are non-ideal, the theorem has not yet been proven
since the proof always requires some arbitrary and {\em ad hoc}
supplementary assumptions.
In this paper, we state a  new and rather weak 
supplementary assumption for the ensemble of
photons that emerge from the polarizers, and 
we show that the conjunction of Einstein's locality 
with this
assumption leads to validity of an inequality that
is violated by a factor as large as $1.5$ in the case of real
experiments.
Moreover, the present supplementary assumption is considerably
weaker and more general than 
Clauser, Horne, Shimony, Holt supplementary assumption.
\end {abstract}
\pagebreak

Einstein, Podolsky, and Rosen (EPR) \cite{1} theorem of $1935$ is a 
proof that if local realism holds, then
quantum mechanics must be an incomplete theory.
In 1965, Bell \cite{2} showed that the assumption of local realism,
as postulated by EPR,
implies some
constraints on the statistics of two spatially separated particles.
These constraints which are collectively known as Bell inequalities
are sometimes grossly violated by 
quantum mechanics
in the case of ideal 
experiments.
Bell's \cite{2} theorem is therefore a proof that local interpretations
of quantum 
mechanics are impossible. 

Bell's original argument, however, can not be experimentally 
tested because it relies on ideal polarizers and
detectors.
Faced with this problem,
correlation inequalities have been derived for non-ideal systems
[3-10].
However, quantum mechanics does not violate any
of these inequalities. 
In the case of real experiments,
the violation
of these inequalities  arises only when
some {\em ad hoc}
assumptions, whose correctness can never be tested,
are supplemented to the original inequalities.

In this paper, we state a new and rather weak
supplementary assumption for
the ensemble of photons that emerge from the polarizers and 
we show this assumption is sufficient to make experiments 
which are feasible with present technology applicable as a test 
of locality. In particular, we deduce a 
correlation inequality
for two-channel polarizer systems 
and we show that 
quantum mechanics violates this inequality 
by a maximum factor of $1.5$.
Since quantum mechanics violates 
the previous inequalities [4-10] 
by a maximum factor of
$\sqrt 2$,
the magnitude of
violation of the present inequality 
is approximately $20.7\%$ larger than that of previous 
inequalities.

We start by considering the Bohm's \cite{12} version of EPR experiment
in which an unstable source emits pairs of photons in
a cascade from state $J=1$ to $J=0$.
The source is viewed by two
apparatuses.
The first (second) apparatus consists of a polarizer
$P_1 \left(P_2 \right)$
set at angle $\mbox{\boldmath $m$} \left(
\mbox{\boldmath $n$} \right)$, and
two detectors
$S_{1}^{\,\pm} \left (S_{2}^{\,\pm} \right)$
put along the ordinary and the extraordinary beams.
We consider a particular photon that passes through the 
first [second] polarizer,
and assign
observable $A(\mbox{\boldmath $m$})$
[$B(\mbox{\boldmath $n$})$] to this photon. These 
observables can take the following values:
\begin{eqnarray}
A(\mbox{\boldmath $m$})=\left\{\begin{array}{cl}
1, & \, \,  \mbox{the photon emerges along the ordinary axis}\\
0, & \, \, \mbox{the photon is absorbed by the polarizer}\\
-1, & \, \,  \mbox{the photon emerges along the extra-ordinary axis}\\
\end{array}
\right.
\end{eqnarray}
and
\begin{eqnarray}
B(\mbox{\boldmath $n$})=\left\{\begin{array}{cl}
1, & \, \,  \mbox{the photon emerges along the ordinary axis}\\
0, & \, \, \mbox{the photon is absorbed by the polarizer}\\
-1, & \, \,  \mbox{the photon emerges along the extra-ordinary axis}.\\
\end{array}
\right.
\end{eqnarray}

During a period of time $T$ 
while the polarizers are set along axes
$\mbox{\boldmath $m$}$ and 
$\mbox{\boldmath $n$}$, the source emits, say, $N$ pairs of
photons.
Let $N^{\,+\,+}\left(\mbox{\boldmath $m,n$}\right)$
[$N^{\,-\,-}\left(\mbox{\boldmath $m,n$}\right)$]
be the number of photon pairs that emerge along the 
ordinary [extra-ordinary] axes; 
$N^{\,+\,-}\left(\mbox{\boldmath $m,n$}\right)$
[$N^{\,-\,+}\left(\mbox{\boldmath $m,n$}\right)$]
be the number of photon pairs in which the first
photon emerges
along the 
ordinary [extra-ordinary] axis
and the second
photon emerges
along the 
extra-ordinary (ordinary) axis;
$N^{+\,0}\left(\mbox{\boldmath $m,n$}\right)$
[$N^{-\,0}\left(\mbox{\boldmath $m,n$}\right)$]
the number of pairs in which the first photon emerges along
the ordinary [extra-ordinary]
axis and the second photon is absorbed by the polarizer;
$N^{0\,+}\left(\mbox{\boldmath $m,n$}\right)$
[$N^{0\,-}\left(\mbox{\boldmath $m,n$}\right)$]
the number of pairs in which the first photon
is absorbed by the polarizer and the second photons
emerges along
the ordinary [extra-ordinary] axis, and finally let
$N^{\,0\,0}\left(\mbox{\boldmath $m,n$}\right)$ be
the number of photon pairs that are emitted by the source
 but absorbed by the polarizers.
If the time $T$ is sufficiently
long, then the ensemble probabilities
are defined as
\begin{eqnarray}{\nonumber}
&&p^{\;\pm\;\pm} \left(\mbox{\boldmath $m,n$} \right)=
\frac{N^{\;\pm\;\pm} \left(\mbox{\boldmath $m,n$} \right)}{N}, \qquad
p^{\;\pm\; 0} \left(\mbox{\boldmath $m,n$} \right)=
\frac{N^{\;\pm \;  0} \left(\mbox{\boldmath $m,n$} \right)}{N}, \\
&&p^{\;0\;\pm} \left(\mbox{\boldmath $m,n$} \right)=
\frac{N^{\;0\;\pm} \left(\mbox{\boldmath $m,n$} \right)}{N}, \qquad
p^{\;0\; 0} \left(\mbox{\boldmath $m,n$} \right)=
\frac{N^{\;0\;0} \left(\mbox{\boldmath $m,n$} \right)}{N}.
\end{eqnarray}
In terms of the ensemble probabilities, the expected value
$e \,\left (\mbox{\boldmath $a,b$} \right)$
is defined as
\begin{eqnarray} 
e \,\left (\mbox{\boldmath $a,b$} \right) = 
p^{+\, +} \,\left (\mbox{\boldmath $a,b$} \right)-
p^{+\, -} \,\left (\mbox{\boldmath $a,b$} \right)
-p^{-\, +} \,\left (\mbox{\boldmath $a,b$} \right)+
p^{-\, -} \,\left (\mbox{\boldmath $a,b$} \right).
\end{eqnarray}

Now let
$N^{\,+}\left(\mbox{\boldmath $m$}\right)$ 
[$N^{\,+}\left(\mbox{\boldmath $n$}\right)$] be
the number of photons that pass through polarizer $P_1$ [$P_2$]
and emerge along the ordinary axis;
$N^{\,-}\left(\mbox{\boldmath $m$}\right)$ 
[$N^{\,-}\left(\mbox{\boldmath $n$}\right)$] 
the number of photons that pass through polarizer $P_1$ [$P_2$]
and emerge along the extra-ordinary axis, and
$N^{\,0}\left(\mbox{\boldmath $m$}\right)$
[$N^{\,0}\left(\mbox{\boldmath $n$}\right)$]
the number of photons that are emitted by the source but absorbed by
polarizers $P_1$ [$P_2$].
Again if the time $T$ is sufficiently
long, then the ensemble probabilities 
are defined as
\begin{eqnarray}{\nonumber}
&&p^{\;\pm}(\mbox{\boldmath $m$})=
\frac{N^{\;\pm}(\mbox{\boldmath $m$})}{N}, \qquad
p^{\;0}(\mbox{\boldmath $m$})=
\frac{N^{\;0}(\mbox{\boldmath $m$})}{N}, \\ 
&&p^{\;\pm}(\mbox{\boldmath $n$})=
\frac{N^{\;\pm}(\mbox{\boldmath $n$})}{N}, \qquad
p^{\;0}(\mbox{\boldmath $n$})=
\frac{N^{\;0}(\mbox{\boldmath $n$})}{N}.
\end{eqnarray}
Since
$p^{\;\pm \; 0} \left(\mbox{\boldmath $m,n$} \right)$,
$p^{\;0 \; \pm} \left(\mbox{\boldmath $m,n$} \right)$,
$p^{\;0 \; 0} \left(\mbox{\boldmath $m,n$} \right)$,
$p^{\;0}(\mbox{\boldmath $m$})$, and
$p^{\;0}(\mbox{\boldmath $n$})$
can not be measured in actual experiments, it is crucial that they do
not appear in any
correlation inequality that is used to test locality
(it is important to emphasize that in real experiments, 
due to imperfection of
polarizers and detectors,
$p^{\;\pm \; 0} \left(\mbox{\boldmath $m,n$} \right)$,
$p^{\;0 \; \pm} \left(\mbox{\boldmath $m,n$} \right)$, and
$p^{\;0 \; 0} \left(\mbox{\boldmath $m,n$} \right)$
are non-zero).

Note that Eqs. $3$ and $5$ are quite general and follow from the 
standard rules of probability theory. No assusmption has yet been made
that is not satisfied by quantum mechanics. 
Hereafter, we
focus our attention only on those theories that satisfy
EPR criterion of locality: ``Since at the time of measurement the
two systems no longer interact, no real change can take place in the
second system in consequence of anything that may be done to first
system
\cite {1}''.
Assuming the first polarizer $P_1$ may be set along the axes 
$\mbox{\boldmath $a$}$ or
$\mbox{\boldmath $b$}$
and the second polarizer $P_2$ may also be set along the axes 
$\mbox{\boldmath $a'$}$ or
$\mbox{\boldmath $b'$}$,
EPR's criterion of locality can be translated into
the following relation:
\begin{center}
{\em Locality $\Longrightarrow$
There exists a four-axis probability 
\\
distribution
function 
$p \left( \mbox{\boldmath $a$},\mbox{\boldmath $b$},
\mbox{\boldmath $a'$},\mbox{\boldmath $b'$} \right)$.
\hspace{0.75 in}
(5)}
\end{center}
Relation (5) 
is a very general form of locality that accounts
for correlations subject only to the requirement that
emergence of the first photon through the first polarizer
does not depend on
the orientation of the second polarizer. This assumption
is quite natural since the two photons are spatially separated so that
the orientation of the second polarizer should not influence the
measurement carried out on the first photon.
The extreme generality of relation 5 as the requirement for 
locality has been discussed in detail by Wigner 
\cite{11} [see also Selleri in
\cite{3}].

In the following we show that 
relation $(5)$ leads to validity of an equality
that is sometimes grossly violated by
the quantum mechanical predictions.
First we need to prove the following algebraic theorem.
\\
{\it Theorem:} If local realism holds, that is, if
the four-axis probability distribution function 
$p\,(\mbox{\boldmath $a$},\mbox{\boldmath $b$},
\mbox{\boldmath $a'$},\mbox{\boldmath $b'$})$ 
exists,
then the following inequality always holds:
\begin{eqnarray} {\nonumber}
&&e\, (\mbox{\boldmath $a,\, b$}) +
e\,(\mbox{\boldmath $b',\, a$})
+e\,(\mbox{\boldmath $b,\, a'$})
\ge
2p^{+ +}(\mbox{\boldmath $a',\, b'$}) +
2p^{- \,-}(\mbox{\boldmath $a',\, b'$}) \\
&&-p^{+}(\mbox{\boldmath $a'$})
-p^{-}(\mbox{\boldmath $a'$})
-p^{+}(\mbox{\boldmath $b'$})
-p^{-}(\mbox{\boldmath $b'$})
-1,
\end{eqnarray}
{\it Proof:}
Assuming 
$p \, (\mbox{\boldmath $a$},\mbox{\boldmath $b$},
\mbox{\boldmath $a'$},\mbox{\boldmath $b'$})$ exists, the
LHS of (6) is defined as
\begin{eqnarray} 
e\,(\mbox{\boldmath $a,\, b$}) +
e\,(\mbox{\boldmath $b',\, a$})
+e\,(\mbox{\boldmath $b,\, a'$})&=& 
\sum_{i\, j} p^{i\,j}(\mbox{\boldmath $a$},\mbox{\boldmath $b$},
\mbox{\boldmath $a'$},\mbox{\boldmath $b'$}) \\ \nonumber
&&[A(\mbox{\boldmath $a$})
B(\mbox{\boldmath $b$})+
A(\mbox{\boldmath $a$})
B(\mbox{\boldmath $b'$})+
A(\mbox{\boldmath $a'$})
B(\mbox{\boldmath $b$})] 
\end{eqnarray}
where $i$ and $j$ belong to the set $\{+,\, -,\,0\}$.
We now consider the following $9$ cases:
\\
(i) First assume $A(\mbox{\boldmath $a'$})=1$ and
$B(\mbox{\boldmath $b'$})=1$, then
\\
$A(\mbox{\boldmath $a$})
B(\mbox{\boldmath $b$})+
A(\mbox{\boldmath $a$})
B(\mbox{\boldmath $b'$})+
A(\mbox{\boldmath $a'$})
B(\mbox{\boldmath $b$}) \ge -1$.
\\
(ii) Next assume $A(\mbox{\boldmath $a'$})=-1$ and
$B(\mbox{\boldmath $b'$})=-1$, then 
\\
$A(\mbox{\boldmath $a$})
B(\mbox{\boldmath $b$})+
A(\mbox{\boldmath $a$})
B(\mbox{\boldmath $b'$})+
A(\mbox{\boldmath $a'$})
B(\mbox{\boldmath $b$})  \ge -1$.
\\
(iii) Next assume $A(\mbox{\boldmath $a'$})=1$ and
$B(\mbox{\boldmath $b'$})=-1$, then
\\
$A(\mbox{\boldmath $a$})
B(\mbox{\boldmath $b$})+
A(\mbox{\boldmath $a$})
B(\mbox{\boldmath $b'$})+
A(\mbox{\boldmath $a'$})
B(\mbox{\boldmath $b$}) \ge -3$.
\\
(iv) Next assume $A(\mbox{\boldmath $a'$})=-1$ and
$B(\mbox{\boldmath $b'$})=1$, then
\\
$A(\mbox{\boldmath $a$})
B(\mbox{\boldmath $b$})+
A(\mbox{\boldmath $a$})
B(\mbox{\boldmath $b'$})+
A(\mbox{\boldmath $a'$})
B(\mbox{\boldmath $b$}) \ge -3$.
\\
(v) Next assume $A(\mbox{\boldmath $a'$})=1$ and
$B(\mbox{\boldmath $b'$})=0$, then
\\
$A(\mbox{\boldmath $a$})
B(\mbox{\boldmath $b$})+
A(\mbox{\boldmath $a$})
B(\mbox{\boldmath $b'$})+
A(\mbox{\boldmath $a'$})
B(\mbox{\boldmath $b$}) \ge -2$.
\\
(vi) Next assume $A(\mbox{\boldmath $a'$})=-1$ and
$B(\mbox{\boldmath $b'$})=0$, then 
\\
$A(\mbox{\boldmath $a$})
B(\mbox{\boldmath $b$})+
A(\mbox{\boldmath $a$})
B(\mbox{\boldmath $b'$})+
A(\mbox{\boldmath $a'$})
B(\mbox{\boldmath $b$}) \ge  -2$.
\\
(vii) Next assume $A(\mbox{\boldmath $a'$})=0$ and
$B(\mbox{\boldmath $b'$})=1$, then
\\
$A(\mbox{\boldmath $a$})
B(\mbox{\boldmath $b$})+
A(\mbox{\boldmath $a$})
B(\mbox{\boldmath $b'$})+
A(\mbox{\boldmath $a'$})
B(\mbox{\boldmath $b$}) \ge -2$.
\\
(viii) Next assume $A(\mbox{\boldmath $a'$})=0$ and
$B(\mbox{\boldmath $b'$})=-1$, then
\\
$A(\mbox{\boldmath $a$})
B(\mbox{\boldmath $b$})+
A(\mbox{\boldmath $a$})
B(\mbox{\boldmath $b'$})+
A(\mbox{\boldmath $a'$})
B(\mbox{\boldmath $b$}) \ge -2$.
\\
(ix) Finally assume $A(\mbox{\boldmath $a'$})=0$ and
$B(\mbox{\boldmath $b'$})=0$, then 
\\
$A(\mbox{\boldmath $a$})
B(\mbox{\boldmath $b$})+
A(\mbox{\boldmath $a$})
B(\mbox{\boldmath $b'$})+
A(\mbox{\boldmath $a'$})
B(\mbox{\boldmath $b$}) \ge -1$.
\\
Using relations (i-ix), we obtain the following
upper bound on $(7)$.
\begin{eqnarray} \nonumber 
&&e\,(\mbox{\boldmath $a,\, b$}) +
e\,(\mbox{\boldmath $b',\, a$})
+e\,(\mbox{\boldmath $b,\, a'$}) \ge
-3 p^{+ \, -} (\mbox{\boldmath $a'$},\mbox{\boldmath $b'$})
-3 p^{- \, +} (\mbox{\boldmath $a'$},\mbox{\boldmath $b'$}) \\ \nonumber
&&-2 p^{+ \, 0} (\mbox{\boldmath $a'$},\mbox{\boldmath $b'$})
-2 p^{0 \, +} (\mbox{\boldmath $a'$},\mbox{\boldmath $b'$})
-2 p^{- \, 0} (\mbox{\boldmath $a'$},\mbox{\boldmath $b'$})
-2 p^{0 \, -} (\mbox{\boldmath $a'$},\mbox{\boldmath $b'$}) 
\\ 
&&- p^{0 \, 0} (\mbox{\boldmath $a'$},\mbox{\boldmath $b'$})
- p^{+ \, +} (\mbox{\boldmath $a'$},\mbox{\boldmath $b'$})
- p^{- \, -} (\mbox{\boldmath $a'$},\mbox{\boldmath $b'$})
\end{eqnarray} 
We now note that
\begin{eqnarray} \nonumber
&&p^{+ \,+} (\mbox{\boldmath $a',\, b'$})+
p^{+ \,-} (\mbox{\boldmath $a',\, b'$})+
p^{- \,+} (\mbox{\boldmath $a',\, b'$})+
p^{- \,-} (\mbox{\boldmath $a',\, b'$})+ \\ \nonumber
&&p^{+ \,0} (\mbox{\boldmath $a',\, b'$})+
p^{0 \,+} (\mbox{\boldmath $a',\, b'$})+
p^{0 \,0} (\mbox{\boldmath $a',\, b'$})=1, \\ \nonumber
&&p^{+}(\mbox{\boldmath $a'$})=
p^{+ \,+} (\mbox{\boldmath $a',\, b'$})+
p^{+ \,-} (\mbox{\boldmath $a',\, b'$})+
p^{+ \,0} (\mbox{\boldmath $a',\, b'$}), \\ \nonumber
&&p^{-}(\mbox{\boldmath $a'$})= 
p^{- \,+} (\mbox{\boldmath $a',\, b'$})+
p^{- \,-} (\mbox{\boldmath $a',\, b'$})+
p^{- \,0} (\mbox{\boldmath $a',\, b'$}), \\ \nonumber
&&p^{+}(\mbox{\boldmath $b'$})=
p^{+ \,+} (\mbox{\boldmath $a',\, b'$})+
p^{- \,+} (\mbox{\boldmath $a',\, b'$})+
p^{0 \,+} (\mbox{\boldmath $a',\, b'$}), \\
&&p^{-}(\mbox{\boldmath $b'$})= 
p^{+ \,-} (\mbox{\boldmath $a',\, b'$})+
p^{- \,-} (\mbox{\boldmath $a',\, b'$})+
p^{0 \,-} (\mbox{\boldmath $a',\, b'$}).
\end{eqnarray}
Substituting $(9)$ in $(8)$ and rearranging, we obtain
\begin{eqnarray} {\nonumber}
&&e\,(\mbox{\boldmath $a,\, b$}) +
e\,(\mbox{\boldmath $b',\, a$})
+e\,(\mbox{\boldmath $b,\, a'$})
-2p^{+ +}(\mbox{\boldmath $a',\, b'$}) 
-2p^{- \,-}(\mbox{\boldmath $a',\, b'$}) \\
&&+p^{+}(\mbox{\boldmath $a'$})
+p^{-}(\mbox{\boldmath $a'$})
+p^{+}(\mbox{\boldmath $b'$})
+p^{-}(\mbox{\boldmath $b'$})
\ge
-1,
\end{eqnarray}
and the theorem is proved.

First we consider an
atomic cascade experiment in which
polarizers and detectors are ideal.
In an ideal experiment while the polarizers are set along
arbitrary axes $\mbox{\boldmath $m$}$ and
$\mbox{\boldmath $n$}$, all emitted photons pass through the 
polarizers and are analyzed. Thus
the probability that a
photon is absorbed by the polarizer or is not analyzed is zero, i.e.,
\begin{eqnarray}
p^{0} \,\left ( \mbox{\boldmath $m$} \right)=
p^{0} \,\left ( \mbox{\boldmath $n$} \right)=
p^{\pm\, 0} \,\left ( \mbox{\boldmath $m, n$} \right)=
p^{0\, \pm} \,\left ( \mbox{\boldmath $m, n$} \right)=
p^{0\, 0} \,\left ( \mbox{\boldmath $m, n$} \right)=0.
\end{eqnarray}

Inequality $(10)$
may be considerably simplified if we invoke some
of the symmetries that are exhibited in atomic-cascade photon
experiments. For a pair of photons in
cascade from state $J=1$ to $J=0$, the
quantum mechanical detection probabilities $p^{\pm\, \pm}_{QM}$ and
expected value $e_{QM}$ exhibit the following
symmetry
\begin{eqnarray}
p^{\pm\, \pm}_{QM} \,\left (\mbox{\boldmath $a,b$} \right)=
p^{\pm\, \pm}_{QM}
\,\left ( \mid\mbox{\boldmath $a-b$} \mid\right), \qquad
e_{QM} \,\left (\mbox{\boldmath $a,b$} \right)=
e_{QM} \,\left ( \mid\mbox{\boldmath $a-b$} \mid\right).
\end{eqnarray}
We assume that the
local theories also exhibit the same symmetry
\begin{eqnarray}
p^{\pm\, \pm} \,\left (\mbox{\boldmath $a,b$} \right)=
p^{\pm\, \pm} \,\left ( \mid\mbox{\boldmath $a-b$} \mid\right), \qquad
e \,\left (\mbox{\boldmath $a,b$} \right)=
e \,\left ( \mid\mbox{\boldmath $a-b$} \mid\right).
\end{eqnarray}

Now if we choose the following orientation
$\left ( \mbox{\boldmath $a , \, b$} \right)=
\left( \mbox{\boldmath $b' , \, a$} \right) =
\left( \mbox{\boldmath $b , \, a'$}\right) = 120^\circ $
and
$\left( \mbox {\boldmath $a', \, b'$} \right) =0^\circ $, and 
using $(13)$,
inequality $(10)$
becomes
\begin{eqnarray} \nonumber
&&3 e \left (120^\circ \right)
-2 p^{+\, +} \left (0^\circ \right)
-2 p^{-\, -} \left (0^\circ \right)+\\
&&p^{+} (\mbox{\boldmath $a'$})+
p^{-} (\mbox{\boldmath $a'$})+
p^{+} (\mbox{\boldmath $b'$})+
p^{-} (\mbox{\boldmath $b'$}) \ge -1.
\end{eqnarray}
For ideal polarizers and detectors,
the single and joint detection
probabilities for a pair of photons in
a cascade from state $J=1$ to $J=0$
are given by
\begin{eqnarray} {\nonumber}
&&e \left ( \mbox{\boldmath $m,\, n$} \right)=
e \left(\theta \right) = \cos 2 \theta, \qquad
p^{+} (\mbox{\boldmath $a'$})=
p^{-} (\mbox{\boldmath $a'$})=
p^{+} (\mbox{\boldmath $b'$})=
p^{-} (\mbox{\boldmath $b'$})= \frac{1}{2}, \\ \nonumber
&&p^{+\, +} \left ( \mbox{\boldmath $m,\, n$} \right)=
p^{+\, +}\left(\theta \right) =
\frac{\cos^2 \theta}{2},
\qquad
p^{-\, -} \left ( \mbox{\boldmath $m,\, n$} \right)=
p^{-\, -} \left(\theta \right)
=\frac{\cos^2 \theta}{2}.
\\
\end{eqnarray}
Substituting $(15)$ in $(14)$, we obtain
\begin{eqnarray} {\nonumber}
&&3 \cos \left (240^\circ \right)
 -2 \frac{\cos^2 \left (0^\circ \right)}
{2}
-2 \frac{\cos^2 \left (0^\circ \right)}{2} + \frac{1}{2}
 + \frac{1}{2}
 + \frac{1}{2}
 + \frac{1}{2}\\
&&=3*(-0.5)-2*\frac{1}{2}-2*\frac{1}{2}+2 \ge -1,
\end{eqnarray}
or
\begin{eqnarray}
-1.5 \ge -1,
\end{eqnarray}
which violates inequality $(10)$
by a factor of $1.5$
in the case of ideal experiments.

It is important to emphasize that in the case of ideal experiments,
the present
inequality immediately
reduces to
Bell's original inequality of 1965 \cite{2}.
To show this,
we assume
$\mbox{\boldmath $a'$}$ and $\mbox{\boldmath $b'$}$
are along the same direction, using $(15)$, we have
$p^{+\, +}(\mbox{\boldmath $a',\, b'$})=
p^{-\, -}(\mbox{\boldmath $a',\, b'$})=
\frac{\displaystyle 1}{\displaystyle 2}$,
$p^{+} \,\left (\mbox{\boldmath $a'$} \right)=
p^{-} \,\left (\mbox{\boldmath $a'$} \right)=
p^{+} \,\left (\mbox{\boldmath $b'$} \right)=
p^{-} \,\left (\mbox{\boldmath $b'$} \right)=
\frac{\displaystyle 1}{\displaystyle 2}$.
Inequality $(10)$ therefore
becomes
\begin{eqnarray}
e\,(\mbox{\boldmath $a,\, b$})+
e\,(\mbox{\boldmath $b',\, a$})+
e\,(\mbox{\boldmath $a',\, b$})
\ge -1,
\end{eqnarray}
which is the same as Bell's original inequality of $1965$.

We have thus shown that in an ideal experiment where
$p^{+\, +}(\mbox{\boldmath $m,\, n$})+
p^{+\, -}(\mbox{\boldmath $m,\, n$})+
p^{-\, +}(\mbox{\boldmath $m,\, n$})+
p^{+\, -}(\mbox{\boldmath $m,\, n$}) =1$,
Bell's original inequality $(18)$ is sufficient and
there is no need for inequality $(10)$.
However, in a
real experiment where
$p^{0}(\mbox{\boldmath $m$})$,
$p^{0}(\mbox{\boldmath $n$})$,
$p^{\pm\, 0}(\mbox{\boldmath $m,\, n$})$,
$p^{0\, \pm}(\mbox{\boldmath $m,\, n$})$, and
$p^{0\, 0}(\mbox{\boldmath $m,\, n$})$ are non-zero,
i. e., for the case where
$p^{+\, +}(\mbox{\boldmath $m,\, n$})+
p^{+\, -}(\mbox{\boldmath $m,\, n$})+
p^{-\, +}(\mbox{\boldmath $m,\, n$})+
p^{+\, -}(\mbox{\boldmath $m,\, n$}) <1$,
inequality $(10)$  is a distinct and new inequality.

We now consider a real atomic cascade 
experiment in which
polarizers and detectors are non-ideal.
In the cascade experiment 
an atom emits two photons in
a cascade from state $J=1$ to $J=0$. 
Since the pair of photons
have zero angular momentum, they propagate in the form of spherical
wave. Thus 
the
joint probability for single transmission and single detection
are given by
\begin{eqnarray} 
D^{+} \left ( \mbox{\boldmath $a$} \right )=
D^{-} \left ( \mbox{\boldmath $a$} \right )=
\eta \left ({\frac{\displaystyle \Omega}{\displaystyle 8 \pi}}
\right), \qquad
D^{+} \left ( \mbox{\boldmath $b$} \right )=
D^{-} \left ( \mbox{\boldmath $b$} \right )=
\eta \left ({\frac{\displaystyle \Omega}{\displaystyle 8 \pi}}
\right).
\end{eqnarray} 
Similarly the joint probability 
for double transmission and double detection
are given by
\begin{eqnarray}  \nonumber
D^{+ \, +} \left ( \mbox{\boldmath $a,\, b$} \right )=
D^{- \, -} \left ( \mbox{\boldmath $a,\, b$} \right )=
\eta^2 \left ({\frac{\displaystyle \Omega}{\displaystyle 8 \pi}}
\right)^2
g \left (\theta,\phi \right )
\left[1+F \left (\theta,\phi \right )
\cos 2 \left ( \mbox{\boldmath $a- b$} \right ) \right ], \\  \nonumber
D^{+ \, -} \left ( \mbox{\boldmath $a,\, b$} \right )=
D^{- \, +} \left ( \mbox{\boldmath $a,\, b$} \right )=
\eta^2 \left ({\frac{\displaystyle \Omega}
{\displaystyle 8 \pi}}\right)^2
g \left (\theta,\phi \right )
\left[1-F \left (\theta,\phi \right )
\cos 2\left ( \mbox{\boldmath $a- b$} \right ) \right],
\\
\end {eqnarray}
where $\eta$ is the quantum efficiency of the detectors,
$\Omega$ is the solid angle of the detector,
$\cos \theta=\mbox{\boldmath $a. b$}$,
and angle $\phi$ is related to $\Omega$ by
\begin{eqnarray}
\Omega=2 \pi \left (1-\cos \phi \right).
\end{eqnarray}
The function
$g \left (\theta,\phi \right )$ is the angular correlation function
and in the special case is given by
\begin{eqnarray}
g \left (\pi, \phi \right ) = 1+
\frac{1}{8} \cos^2 \phi \left (1 + \cos \phi \right)^2.
\end{eqnarray}
The function
$F \left (\theta,\phi \right )$ is the so-called depolarization
factor and for the special case $\theta=\pi$ and small $\phi$
is given by
\begin{eqnarray}
F (\pi, \phi) \approx 1- \frac{2}{3} \left (1-\cos \phi
\right)^{2}.
\end{eqnarray}
The function $F \left (\theta,\phi \right )$,
in general, is very close to $1$.
Finally the expected values for double transmission and double detection
is defined as
\begin{eqnarray} 
E \,\left (\mbox{\boldmath $a,b$} \right) = 
D^{+\, +} \,\left (\mbox{\boldmath $a,b$} \right)-
D^{+\, -} \,\left (\mbox{\boldmath $a,b$} \right)
-D^{-\, +} \,\left (\mbox{\boldmath $a,b$} \right)+
D^{-\, -} \,\left (\mbox{\boldmath $a,b$} \right).
\end{eqnarray}

Note that in an actual experimnet, the measurable quantities are
the joint probabilities for transmission and detection, i.e,
the probabilities
in Eqs. $(19)$ and $(20)$.
However, the probabilities that appear in the inequality
that we derived in this paper,
i.e., inequality $(10)$, are {\em not} joint probabilities
for transmission and detection, rather they are only the
probabilities for transmission through the 
polarizers. One can certainly
attempt to redefine the probabilities in inequality $(10)$
by using the measurable joint probabilities for detection and 
transmission, but the trouble is that in 
experiments which are feasible with present technology [5,13],
because
$\frac{\displaystyle \Omega}{\displaystyle  4 \pi} \ll 1$,
the probabilities
$D^{\pm \, \pm} \left ( \mbox{\boldmath $a,\, b$} \right )$ are
of the order $10^{-2}$ which are far too small to lead to 
violation of Bell's inequality.

We solve this problem by means of the following supplementary
assumption:
Given that an ensemble of photon emerges from two (one) polarizers,
the probability of their
double (single) detection is equal to the sum of joint probabilities 
for double (single) transmission and double (single) detection.
Calling $T_0 \left ( \mbox{\boldmath $m,\, n$} \right )$,
[($t_0\left ( \mbox{\boldmath $m$} \right )$,
($t_0\left ( \mbox{\boldmath $n$} \right )$],
the sum of joint probabilities 
for double [single] transmission and double [single] detection
the above 
supplementary assumption can be translated into the following relation:
\begin{eqnarray} \nonumber
&&D^{\pm \, \pm}(\mbox{\boldmath $m,\, n$})=
T_0 \left ( \mbox{\boldmath $m,\, n$} \right )
p^{\pm \, \pm}(\mbox{\boldmath $m,\, n$}), \\ 
&&D^{\pm }(\mbox{\boldmath $m$})=t_0\left (\mbox{\boldmath $m$} \right)
p^{\pm }(\mbox{\boldmath $m$}), \qquad
D^{\pm }(\mbox{\boldmath $n$})=t_0\left ( \mbox{\boldmath $n$}\right)
p^{\pm }(\mbox{\boldmath $n$}),
\end{eqnarray}
where 
\begin{eqnarray} \nonumber
&&T_0(\mbox{\boldmath $m,\, n$})=
D^{+ \, +}(\mbox{\boldmath $m,\, n$})+
D^{+ \, -}(\mbox{\boldmath $m,\, n$})+
D^{- \, +}(\mbox{\boldmath $m,\, n$})+
D^{- \, -}(\mbox{\boldmath $m,\, n$}),\\ 
&&t_0(\mbox{\boldmath $m$})=
D^{+}(\mbox{\boldmath $m$})+
D^{-}(\mbox{\boldmath $m$}), \qquad
t_0(\mbox{\boldmath $n$})=
D^{+}(\mbox{\boldmath $n$})+
D^{-}(\mbox{\boldmath $n$}).
\end{eqnarray}
Using Eqs. $(19)$ and $(20)$, it can be seen that
the quantum mechanical prediction for  $T_0(\mbox{\boldmath $m,\, n$})$,
$t_0(\mbox{\boldmath $m$})$, and $t_0(\mbox{\boldmath $n$})$ are
\begin{eqnarray}
&&T_0(\mbox{\boldmath $m,\, n$})=
\eta^2 \left ({\frac{\displaystyle \Omega}{\displaystyle 4 \pi}}
\right)^2
g \left (\theta,\phi \right ), \\ \nonumber
&&t_0(\mbox{\boldmath $m$})=
\eta \left ({\frac{\displaystyle \Omega}{\displaystyle 4 \pi}} 
\right), \qquad
t_0(\mbox{\boldmath $n$})=
\eta \left ({\frac{\displaystyle \Omega}{\displaystyle 4 \pi}} 
\right).
\end{eqnarray}
The supplementary assumption $(25)$
allows us to transform Bell's inequality $(10)$ into a measurable
inequality. Substituting in $(25)$ in $(10)$, we obtain
\begin{eqnarray} {\nonumber}
&&\frac{E(\mbox{\boldmath $a,\, b$})}{T_0(\mbox{\boldmath $a,\, b$})} +
\frac{E(\mbox{\boldmath $b',\, a$})}{T_0(\mbox{\boldmath $b',\, a$})}
+\frac{E(\mbox{\boldmath $b,\, a'$})}{T_0(\mbox{\boldmath $b,\, a'$})}
-2\frac{D^{+ +}(\mbox{\boldmath $a',\, b'$})}
{T_0(\mbox{\boldmath $a',\, b'$})}-
2\frac{D^{- \,-}(\mbox{\boldmath $a',\, b'$})}
{T_0(\mbox{\boldmath $a',\, b'$})}\\ 
&&+\frac{D^{+}(\mbox{\boldmath $a'$})}{t_0(\mbox{\boldmath $a'$})}
+\frac{D^{-}(\mbox{\boldmath $a'$})}{t_0(\mbox{\boldmath $a'$})}
+\frac{D^{+}(\mbox{\boldmath $b'$})}{t_0(\mbox{\boldmath $b'$})}
+\frac{D^{-}(\mbox{\boldmath $b'$})}{t_0(\mbox{\boldmath $b'$})}
\ge
-1.
\end{eqnarray} 
\noindent Note that 
the number of emissions $N$ from the source
is eliminated from the ratio 
of in inequality $(28)$.
Using the quantum mechanical predictions
$(19)$ and $(20)$, it can easily be seen that
quantum mechanics violates
inequality $(28)$
in the case of real experiments where the solid angle covered
by the aperture of the apparatus, $\Omega$, is  much less than
$4 \pi$. In particular, the magnitude of violation is maximized if the
following orientations are chosen
$\left (\mbox{\boldmath $a , \, b$} \right)=
\left( \mbox{\boldmath $b' , \, a$} \right)=
\left( \mbox{\boldmath $b , \, a'$} \right)= 120^\circ $
and
$ \left( \mbox {\boldmath $a', \, b'$} \right)= 0^\circ $.
Using the quantum mechanical probabilities
[i.e., Eqs. $(19)$ and $(20)$],
inequality $(28)$ becomes
$-1.5 \geq -1$,
i.e., quantum mechanics violates 
inequality $(28)$ by a factor of $1.5$ 
in the case of real experiments
(here for simplicity, we have assumed
$F \left (\theta,\phi \right ) =1$; this is a good approximation
even in the case of real experiments. In
actual experiments where the solid angle of detectors
$\phi$ is usually less than $\pi/6$, from $(23)$ it can be seen that
$F (\theta, \pi/6) \approx 0.99$).

Inequality $(28)$
may be considerably simplified if we invoke some
of the symmetries that are exhibited in atomic-cascade photon
experiments. For a pair of photons in
cascade from state $J=1$ to $J=0$, the
quantum mechanical detection probabilities $p^{\pm\, \pm}_{QM}$ and
expected value $E_{QM}$ exhibit the following
symmetry
\begin{eqnarray}
D^{\pm\, \pm}_{QM} \,\left (\mbox{\boldmath $a,b$} \right)=
D^{\pm\, \pm}_{QM}
\,\left ( \mid\mbox{\boldmath $a-b$} \mid\right), \qquad
E_{QM} \,\left (\mbox{\boldmath $a,b$} \right)=
E_{QM} \,\left ( \mid\mbox{\boldmath $a-b$} \mid\right).
\end{eqnarray}
We assume that the
local theories also exhibit the same symmetry
\begin{eqnarray}
D^{\pm\, \pm} \,\left (\mbox{\boldmath $a,b$} \right)=
D^{\pm\, \pm} \,\left ( \mid\mbox{\boldmath $a-b$} \mid\right), \qquad
E \,\left (\mbox{\boldmath $a,b$} \right)=
E \,\left ( \mid\mbox{\boldmath $a-b$} \mid\right).
\end{eqnarray}
Note that there is no harm in assuming Eq. $(30)$ since it is
subject to experimental test.
We now take $\mbox{\boldmath $a'$}$ and
$\mbox{\boldmath $b'$}$ to be along the same direction,
and we take
$\mbox{\boldmath $a$}$,
$\mbox{\boldmath $b$}$, and $\mbox{\boldmath $a'$}$ to be
three coplanar axes, each making $120^\circ$ with the other
two,
that is we choose the
the following orientations,
$\mid \mbox{\boldmath $a - b$} \mid=
 \mid \mbox{\boldmath $b' - a$} \mid=
 \mid \mbox{\boldmath $b - a'$} \mid= 120^\circ $
and
$ \mid \mbox {\boldmath $a'- b'$} \mid=0^\circ $,
then inequality $(28)$ is simplified to
\begin{eqnarray}
\frac {3E \left( 120^\circ \right)}{T_0\left( 120^\circ \right)}
-\frac{2D^{+\, +}\left( 0^\circ \right)}
{T_0\left( 0^\circ \right)}
-\frac{2D^{- \,-}\left( 0^\circ \right)}{T_0\left( 0^\circ \right)}
+\frac{2D^{+}\left( 0^\circ \right)}{t_0\left( 0^\circ \right)}
+\frac{2D^{-}\left( 0^\circ \right)}{t_0\left( 0^\circ \right)}
\geq -1.
\end{eqnarray}
Again using the quantum mechanical probabilities
[i.e., Eqs. $(19)$ and $(20)$],
inequality $(31)$ becomes
$-1.5 \geq -1$ in the case of real experiments, i.e., 
Quantum mechanics violates inequality  $(31)$
by a factor of 1.5,
whereas it violates previous inequalities
by a factor of $\sqrt 2$.
Thus the magnitude of violation of inequality $(31)$
is approximately $20.7\%$ larger than the magnitude of violation of
the previous inequalities [4-10].

It should be noted that the analysis that led to inequality $(31)$
is not limited to atomic-cascade experiments and
can easily be extended to experiments which use
phase-momentum \cite {14}, or high energy
polarized protons or $\gamma$ photons [15-16] to test Bell's limit.

A final comment is in order about the supplementary assumption 
of this paper. The present assumption is considerably weaker
than CHSH assumption. CHSH supplementary assumption requires that
\begin{eqnarray} 
\frac{D^{\pm \, \pm}(\mbox{\boldmath $m,\, n$})}
{p^{\pm \, \pm}(\mbox{\boldmath $m,\, n$})}=
\frac{D^{\pm \, \pm}(\mbox{\boldmath $m',\, n'$})}
{p^{\pm \, \pm}(\mbox{\boldmath $m',\, n'$})},
\end{eqnarray} \nonumber
and
\begin{eqnarray} 
\frac{D^{\pm \, \pm}(\mbox{\boldmath $m,\, n$})}
{p^{\pm \, \pm}(\mbox{\boldmath $m,\, n$})}=
D(\infty,\, \infty),
\end{eqnarray} 
where
$\mbox{\boldmath $m$}$, $\mbox{\boldmath $m'$}$
are arbitrary axes of the first polarizer and
$\mbox{\boldmath $n$}$, $\mbox{\boldmath $n'$}$
are arbitrary axes of the second polarizer, and
$D(\infty,\, \infty)$
is the the probability of detection in the absence of polarizers.
In contrast, the present supplementary assumption
does not make any assertion about the
orientation of the polarizers, i.e., according to the 
assumption of this paper,
$\frac{\displaystyle D^{\pm \, \pm}(\mbox{\boldmath $m,\, n$})}
{\displaystyle  p^{\pm \, \pm}
(\mbox{\boldmath $m,\, n$})}$ can be larger than or 
smaller than or equal to 
$\frac{\displaystyle D^{\pm \, \pm}(\mbox{\boldmath $m',\, n'$})}
{\displaystyle p^{\pm \, \pm}(\mbox{\boldmath $m',\, n'$})}$.
The present supplementary assumption only requires that
\begin{eqnarray} 
\frac{D^{\pm \, \pm}(\mbox{\boldmath $m,\, n$})}
{p^{\pm \, \pm}(\mbox{\boldmath $m,\, n$})}=
T_0(\mbox{\boldmath $m,\, n$}).
\end{eqnarray}
Since the present supplementary assumption is considerably weaker
than CHSH assumption, an experiment based on inequality $(31)$
refutes a larger family of hidden variable theories than an 
experiment based on CHSH inequality.
Finally it is interesting to note that for an ensemble of photons,
the numerical value for $T_0(\mbox{\boldmath $m,\, n$})$ is the 
same as for $D(\infty,\, \infty)$, i.e.,
\begin{eqnarray} 
T_0(\mbox{\boldmath $m,\, n$})=
D(\infty,\, \infty)=
\eta^2 \left ({\frac{\displaystyle \Omega}{\displaystyle 4 \pi}}
\right)^2
g \left (\theta,\phi \right ).
\end{eqnarray}

In summary, we have derived
a correlation inequality [inequality $(10)$] which can be used to test
locality.
In case of ideal
experiments, this inequality is equivalent to Bell's
original inequality of $1965$ \cite{2}.
However, in the case of real experiments where
polarizers and detectors
are not ideal, inequality $(10)$ is a new and distinct inequality.
We have also demonstrated that
the conjunction of Einstein's locality
[relation $(5)$] with a supplementary assumption [Eqs. $(25)$, which
is considerably weaker than CHSH assumption,
leads to validity of inequality $(28)$ [or $(31)$].
Quantum mechanics violates this inequality by a 
maximum factor of $1.5$. Thus the magnitude of violation of
inequality $(28)$ is approximately $20.7\%$ larger than 
the magnitude of violation of previous 
inequalities [4-10].

\pagebreak

\end{document}